\documentclass[%
 reprint,
superscriptaddress,
 amsmath,amssymb,
 aps,
pre,
]{revtex4-1}
\usepackage{graphicx}
\usepackage{epsfig}
\usepackage{amsfonts}
\usepackage{amsmath}
\usepackage{amssymb}
\usepackage{subfigure}
\usepackage{color}

\newcommand{\ct}{\cite}
\newcommand{\bi}{\bibitem}
\newcommand{\be}{\begin{equation}}
\newcommand{\ee}{\end{equation}}
\newcommand{\ba}{\begin{eqnarray}}
\newcommand{\ea}{\end{eqnarray}}

\newcommand{\la}{\lambda}

\begin{document}
\title{Entanglement entropy of a three-spin interacting spin chain with a time-reversal breaking impurity at one boundary}
\author{Tanay Nag}
\affiliation{Max Planck Institute for the Physics of Complex Systems, N\"othnitzer Str. 38, Dresden 01187, Germany}
\author{Atanu Rajak}
\affiliation{Department of Physics, Jack and Pearl Resnick Institute, Bar-Ilan University, Ramat-Gan 52900, Israel}

\begin{abstract}

We investigate the effect of a time-reversal breaking impurity term (of strength $\la_d$) on
both the equilibrium and non-equilibrium critical properties of entanglement 
entropy (EE) in a three-spin interacting transverse Ising model which can be mapped
to a $p$-wave superconducting chain with next-nearest-neighbor 
hopping and interaction. 
Importantly, we find that the logarithmic scaling of the EE 
with block size remains unaffected by the application of the impurity term, although, the 
coefficient (i.e., central charge) varies logarithmically with the impurity strength for a 
lower range of $\la_d$ and eventually saturates with an exponential damping factor ($\sim \exp(-\la_d)$) for the phase 
boundaries shared with the phase containing two Majorana edge modes. On the other hand, it receives 
a linear correction in term of $\la_d$ for an another phase boundary.
Finally, we focus to study the effect of the impurity in the time evolution of the EE for the critical 
quenching case where impurity term is applied only to the final Hamiltonian.
Interestingly, it has been shown that for all the phase boundaries, in contrary to 
the equilibrium case, the saturation value of 
the EE increases logarithmically with the strength of impurity in a certain regime of $\la_d$ 
and finally, for higher values of $\la_d$, it increases very slowly dictated by 
an exponential damping factor. The impurity induced behavior of EE might bear some deep 
underlying connection to thermalization.


\end{abstract}
\maketitle

\section{Introduction}

Study of various quantum information theoretic measures such as fidelity \ct{gu08},
decoherence \ct{quan06,damski11a}, concurrence \ct{wootters01,osterloh02}, quantum discord \ct{ollivier01}
and entanglement entropy (EE) \ct{vidal03} has grabbed immense attention as it connects the quantum information 
science \ct{kitaev09, nag12b, sachdeva14,suzuki15, nag15, nag16, rajak16}, 
statistical physics and condensed matter physics \ct{chakrabarti96, sachdev99, polkovnikov11, dutta15} 
in a concrete way. All of the above quantities are able to capture the ground state singularity and thus
are used as indicators of quantum phase
transition (QPT). 
For example, the EE of a block of length $l$ 
quantified by von Neumann entropy is given by
\be
S(l)=-{\rm Tr}(\rho_l\ln \rho_l),
\label{ee_def}
\ee
where  the reduced density matrix $\rho_l={\rm Tr}_{L-l}(\rho)$ is obtained after tracing over the 
block of length  $L-l$ from the composite system of length $L$ with pure state density matrix $\rho$.
For a one-dimensional homogeneous critical spin chain with open boundary conditions,
the EE scales with the shortest length scale ($l$) of the system as $S= \frac{c}{6}\ln l+\gamma$, where 
$c$ is a universal quantity and given by the central charge of the underlying conformal field theory, 
whereas $\gamma$ is a non-universal constant \ct{vidal03,calabrese04}. 
In the context of 
 disordered spin chain (i.e., inhomogeneous case), for the critical case,
 it has been shown that the EE still varies logarithmically with
 block size but it acquires an effective central charge different from the
 bare central charge derived in the clean 
 limit \ct{refael07}. It has been shown that the effective central charge 
 also appears for the interface defects in a spin chain~\ct{peschel05}.

At the same time, the study of entanglement spectrum and EE
 in quantum many-body systems  
has initiated a plethora of intensive research to characterize a topological system, through the concept of
quantum entanglement~\ct{li08,yao10,fidkowski10,pollmann10,facchi08}. The topological phases are charaterized
by a topological invariant number 
(Chern number or $Z_2$ invariant or zero energy Majorana modes) and 
this phase supports edge modes \ct{hasan10, qi11}.
In the connection of EE and edge state, it is noteworthy that the equilibrium EE
receives a finite contribution from the localized boundary states in addition to the contribution from 
bulk energy spectrum. The finite contribution from boundaries is associated to the non-zero value 
of the Berry phase~\ct{ryu06}.
A non-extensive correction to the area law of EE, named as topological EE, has been proposed as a tool to
  characterize the topological phases of the system\ct{kitaev06}. Interestingly,
  it has been shown for two dimension spin-orbit coupled superconductor that
the derivatives of the EE with respect to model parameters are sharply peaked
 at the point of topological phase transitions \ct{borchmann14}.

In parallel, the behavior of the EE in quantum systems considering a 
non-equilibrium situation has received an enormous amount of attention in recent years 
~\ct{calabrese05,eisler07,calabrese07,eisler08,igloi09,hsu09,cardy11}. 
The upsurge of such studies is motivated by the
experimental demonstration in optical lattice~\ct{bloch08}.
In particular, an out-of-equilibrium one-dimensional Bose gases has been prepared 
experimentally using the combination of a two-dimensional optical lattice and a 
crossed dipole trap~\ct{kinoshita05,kinoshita06}.
It has been shown that a global sudden quench 
leads to an initial linear rise of the EE with time followed by a saturation~\ct{calabrese05}.
Moreover, the dynamics of EE
in the random transverse-field Ising chain
after a sudden critical quench becomes ultraslow and has a double-logarithmic
time dependence~\ct{igloi12}. 
Also, the robustness of the Majorana zero-mode in the infinite time limit following a 
sudden quench of a one-dimensional p-wave superconductor has been investigated by examining 
the one-particle entanglement spectrum~\ct{chung14}.
The quench dynamics in optical lattice with $\rm Rb^{87}$ atoms is experimentally investigated in the 
context of quantum information namely, the signature of Lieb-Robinson bound of 
light-cone-like spreading of correlations is studied \ct{cheneau12}.
 Moreover, optical interferometry is directly used to 
measure entanglement entropy in a quantum many-body system composed of ultracold bosonic atoms in optical lattices
\ct{islam15}.

In recent years, a considerable amount of work has been carried out to investigate 
the topological properties of one-dimensional $p$-wave Majorana chain~\ct{kitaev01,fulga11,sau12,
lutchyn11,degottardi11,degottardi13,thakurathi13,wdegottardi13,alicea12,rajak14,rajak14b}. 
Our main aim here is to study the effect of a single impurity, located at one of the boundaries,
on the critical behavior of the EE in this model with an additional next nearest neighbor hopping term. 
This single impurity indeed breaks the time-reversal invariance of the system.
We show that in the equilibrium case, the derivative of EE can be used as an indicator of QPTs. 
We find that the scaling relation
of the EE with the subsystem size remains same as obtained in the inhomogeneous case with an effective central charge. 
{This effective central charge here shows a logarithmic scaling relation with $\la_d$ 
for an initial  window of $\la_d$ and eventually saturates for large value of $\la_d$. This phenomena 
is observed at the phase boundaries shared with the phase containing two Majorana modes sitting at each end of the chain. 
On the other hand, for the phase boundary separating topological phase with one Majorana mode from non-topological one, 
the effective central charge acquires a linear correction  due to this impurity term.
Additionally, for the non-equilibrium time evolution of  EE
obtained by adding a boundary impurity term to the critical quenched chain,
 irrespective of the nature of phase boundaries
the logarithmic behavior  and the subsequent exponential scaling show up  only in the saturation value of the EE 
but not in the initial rise of EE.

This paper is organized as follows: In Sec.~\ref{model} we introduce the three-spin interacting 
transverse field Ising model and discuss its phase diagram. We also mention 
the effect of an impurity term on the different phases of the model.
In Sec.~\ref{EE}, we present the method to compute  EE and extend it
to calculate the time-evolution of the EE numerically.
In Sec.~\ref{result}, we illustrate our results for equilibrium as well as for non-equilibrium case.
Finally, we provide our concluding remarks in Sec.~\ref{conclusion}.

\section{Model}
\label{model}
The Hamiltonian we consider here is given by a three-spin interacting transverse Ising model with $N$ spins~~\ct{kopp05}
\be
H~=~-\sum_n(h\sigma_n^z~+~\la_1\sigma_n^x\sigma_{n+1}^x~+~\la_2\sigma_{n-1}^x\sigma_{n}^z\sigma_{n+1}^x),
\label{ham1}
\ee
where $h$, $\la_1$ and $\la_2$ are transverse magnetic field, cooperative interaction and 
three-spin interaction respectively and, $\sigma^{\alpha}$ ($\alpha=x,y,z$) are the standard 
Pauli matrices. Using Jordan-Wigner transformation the 
model can be written in terms of spinless fermions 
with a next-nearest-neighbor 
hopping and superconducting pairing terms. At the same time, 
one can re-write the Hamiltonian in terms of Majorana fermions with open boundary condition (OBC), 
given by
\be
H=-i\Big[-h\sum_{n=1}^Nb_na_n~+~\la_1\sum_{n=1}^{N-1}b_na_{n+1}~+~\la_2\sum_{n=2}^{N-1}b_{n-1}a_{n+1}\Big].
\label{ham2}
\ee 
where, 
$ a_n = c_n^\dagger+c_n, b_n=-i( c_n^\dagger-c_n)$ are Majorana fermions. 
The system discussed above has time-reversal symmetry ($T^2=1$). 
This transformation $T$ is defined as the complex conjugation of all the objects in the Hamiltonian (\ref{ham2}). 
As a result, $T$ leads to $a_n\rightarrow a_n$ and $b_n\rightarrow-b_n$, and hence,
Eq.~(\ref{ham2}) is invariant under time-reversal ~\ct{wdegottardi13}.

We now briefly discuss the phase diagram of the model with $h=1$ (see Fig.~(\ref{pd})). This model has three phases: (i) the
ferromagnetic phase which is topologically non-trivial hosting one unpaired Majorana at each end, i.e.,
$a_1$ and $b_N$, (ii) the paramagnetic phase, a non-topological phase, with no Majorana edge modes,
and (iii) three-spin dominated topological phase with two unpaired Majorana modes at each edge i.e.,
$a_1$ and $a_2$ are at the left boundary with $b_N$ and $b_{N-1}$ exist at the right boundary. 
The detail of the model and phase diagram are discussed in Refs.\ct{niu12,rajak17}. 

It is noteworthy that a similar kind of model namely, cluster-Ising model, has been studied before
in the context  of locating its QPTs between cluster and antiferromagnetic phases
using geometric~\ct{son11} and multipartite~\ct{giampaolo14} entanglement.
Although, the three-spin dominated phase of the model (\ref{ham1}) is analogous to
the cluster phase of the cluster-Ising model, there are a few differences between these two models: 
(i) model given in Eq.~(\ref{ham1}) contains an extra transverse field term, 
(ii) it has $Z_2^T$ symmetry (an anti-unitary $Z_2$ symmetry, $\sigma_x\rightarrow-\sigma_x$)~\ct{friedman17}, 
whereas the cluster-Ising model is symmetric under is symmetric under $Z_2\times Z_2$, and 
(iii) the dual model of (\ref{ham1}) is a transverse XY spin chain, on the other hand, the cluster-Ising 
model is its self-dual.

\begin{figure}[ht]
\begin{center}
\includegraphics[height=4.8cm]{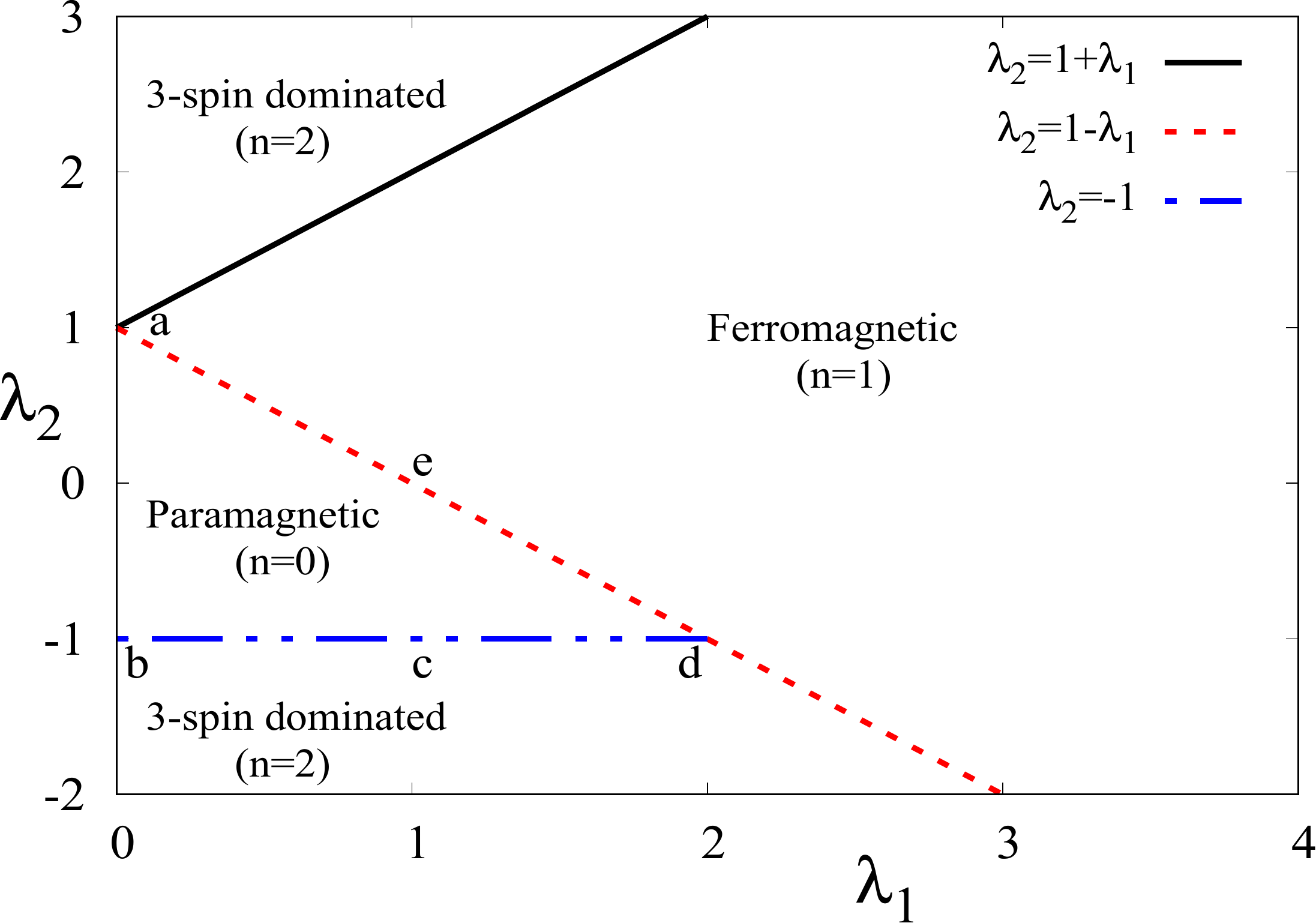}
\end{center}
\caption{(Color online) Zero-temperature phase diagram of the model Hamiltonian (\ref{ham1}) for $h=1$. 
The line $\la_2=1+\la_1$ represents the phase boundary between upper $3$-spin dominated phase (with two zero-energy Majorana 
end modes at each edge) and ferromagnetic phase with one Majorana at each end. There are two more phase 
boundaries, given by $\la_2=1-\la_1$ ($a-e-d$ line) and $\la_2=-1$ ($b-d$ line). The 
paramagnetic region denoted by $n=0$ does not have any zero-energy Majorana mode. In terms of topology 
both the $3$-spin dominated phases have two zero-energy Majorana modes at each end.}
\label{pd}
\end{figure}

Our main goal is to find the effect of an impurity term in the critical behavior of the EE 
for both equilibrium and non-equilibrium cases. To achieve this goal, we introduce an additional term in the 
Hamiltonian~(\ref{ham2}) of the form~\ct{rajak17}
\be
H_{\rm imp}~=~-i\la_da_ja_{q}, q=j+m.
\label{imp_ham}
\ee
The impurity term (\ref{imp_ham}) affects the phase with two Majorana modes at one end (three-spin dominated phase).
We here take into account the impurity term (\ref{imp_ham}), located 
in the left edge of the system, that results in the annihilation of the $a$-type Majorana 
modes, however, the $b$-type Majorana modes residing at the right edge remain intact~\ct{niu12,rajak17}.
In contrast, the phase with one Majorana mode at each end remains unaffected. 
This term breaks the time-reversal symmetry of the system as $T$ reverses the sign of $H_{\rm imp}$.
The annihilation of the edge Majorana mode is related to this symmetry breaking.


In the spin representation, the impurity Hamiltonian is expressed as 
$H_{\rm imp}=\la_d\sigma_j^y\prod_{n=j+1}^{j+m-1}(-\sigma_n^z)\sigma_{j+m}^x$, which
is a string operator. 
Therefore, in the spin language, the breaking of time reversal symmetry of the model in Eq.~(\ref{ham2})
is related to the 
breaking of $Z_2^T$ symmetry as the 
impurity term explicitly breaks this symmetry.
Similarly for the cluster Ising model, we note that if one considers an additional impurity term, breaking
$Z_2\times Z_2$ symmetry of the cluster phase, the existence of the edge Majorana modes 
might get affected in a similar fashion there also.
 
As mentioned already, the impurity term is a non-local string operator,
though, in the Majorana language we can say it is quasi-local (see Eq.~(\ref{imp_ham})).
We hence stress that the impurity is purely quantum
in nature.
In parallel, 
the effect of a classical impurity, the zero
transverse field at the first site of an otherwise homogeneous chain, has been investigated in
a quantum Ising chain by studying the finite size scaling of the magnetizations~\ct{apollaro17}. 
The nature of this 
impurity is classical due to the fact that the left most spin can not flip; in contrary,  the 
impurity term considered in 
 Eq.~(\ref{imp_ham}) is able to flip the spins.
 Simultaneously, transverse Ising model with multi-impurities gives rise to many 
 non-trivial changes in deformation energy and specific heat \ct{huang15}.

Connecting to experimental realization,
the impurity term in Eq.~(\ref{imp_ham}) can be prepared by experiments on entangled atoms in optical lattice. 
The array of parallel spin chains are created from two-dimensional 
degenerate gas of $^{87}{\rm Rb}$ atoms by applying two horizontal 
optical lattice beams. The atoms are initially  prepared in a 
hyperfine state and then impurity is introduced by changing the 
hyperfine structure of one of the atoms~\ct{fukuhara}.

\section{Entanglement entropy}
\label{EE}
We shall here present our numerical method to calculate EE in the Majorana basis under the sudden quenching of a parameter of the chain. 
In order to formulate the non-equilibrium EE, we first briefly discuss the equilibrium EE in the Majorana basis~\ct{latorre04}.
Let us consider a general quadratic form of Eq.~(\ref{ham2}) in terms of Majorana operators
\be
H = \frac{i}{4} \sum_{m,n=1}^{2N} e_m A_{mn} e_n,
\label{eq_ham}
\ee
where $e_{2m-1}=a_m$ and $e_{2m}=b_m$. The matrix elements for $A$ are given by
$A_{n,n+1}=-A_{n+1,n}=1$, $A_{2n,2n+1}=-A_{2n+1,2n}=-\lambda_1$ and 
$A_{2n,2n+3}=-A_{2n+3,2n}=-\lambda_2$. The impurity Hamiltonian~(\ref{imp_ham}) generates two extra 
elements in the matrix $A$: $A_{2j-1,2j+2m-1}=-A_{2j+2m-1,2j-1}=-\lambda_d$.

Let $W \in SO(2N)$ be a special orthogonal matrix that makes $A$ 
block diagonal of the form, 
\be
D = \bigoplus_{k=1}^{N}\tilde{\epsilon}_k \left[
\begin{array}{cc}
0 & 1 \\
-1 &0
\end{array}
\right],
\label{eq_D}
\ee
where $D=WAW^{\dagger}$. Now, a new set of Majorana operators is defined as
\be
d_{p} = \sum_{m=1}^{2N}W_{pm} e_m,\hspace{2mm}{\rm where}\hspace{2mm} p=1,\cdots,2N.
\label{dp_Hi}
\ee
Here, $d_p$ satisfies the relations $d_p^{\dagger} = d_p,~\{d_p,d_q\} = 2\delta_{pq}$
The Hamiltonian in Eq.~(\ref{eq_ham}) in terms of the new Majorana operators is given by
\be
H= \frac{i}{4}\sum_{k=1}^{N} \tilde{\epsilon}_k (d_{2k-1} d_{2k}-d_{2k} d_{2k-1}).
\ee

We can now define the ground state correlation matrix 
$\langle d_p d_q \rangle=\delta_{pq}+ i\Gamma^{B}_{pq}$ where $\Gamma^{B}$ is given by
\be
\Gamma^B = \bigoplus_{k=1}^{N} \left[
\begin{array}{rc}
0 & 1  \\
- 1 &0
\end{array}
\right].
\label{eq_gb}
\ee

Finally, we obtain the correlation matrix $\langle e_m e_n \rangle=\delta_{mn}+ i\Gamma^{A}_{mn}$ in terms of initial Majorana operators $e$. The matrix 
$\Gamma^A$ is found from $\Gamma^B$ using the relation
\be
\Gamma^A=W^{\dagger}\Gamma^BW. 
\label{eq_gammaa}
\ee

As shown in Appendix~\ref{appendixa}, the EE for a block of length $l$ is 
given by
\be
S(l)= -\sum_{n=1}^l \big[\big({1+\eta_n \over 2}\big)\log \big({1+\eta_n \over 2}\big) 
+\big({1-\eta_n \over 2}\big)\log \big({1-\eta_n \over 2}\big)\big],
\label{eq_ee}
\ee
where $\eta_n$ are the imaginary part of the purely imaginary eigenvalues of the 
$2l\times 2l$ skew-symmetric matrix $\Gamma^A$.
We note that the eigenvalues come in pairs.

Now, we shall extend the above equilibrium technique for calculating the EE in a 
situation when the system is suddenly driven out of equilibrium. 
We consider the Majorana Hamiltonian $H$ that instantaneously changes from $H=H_i$ to $H=H_f$ at time $t=0$. 
We therefore have two sets of $W$, namely, $W_i$
and $W_f$ which can transform two 
Hamiltonians $H_i$ and $H_f$ into two block diagonal form as mentioned in Eq. (\ref{eq_D}). 
Similar to Eq.~(\ref{dp_Hi}), we can now define two new set of Majorana operators 
corresponding to the Hamiltonians $H_i$ and $H_f$ given by
\be
d_{p} = \sum_{m=1}^{2N}(W_i)_{pm} e_m \hspace{2mm}{\rm and}\hspace{2mm}d_{p}' = \sum_{m=1}^{2N}(W_f)_{pm} e_m.
\label{dp_Hi_Hf}
\ee

Let us assume that $|\psi_i\rangle$ is the ground state(in the Majorana basis)  of the initial Hamiltonian $H_i$.
Using the relation $\langle\psi_i|e_m e_n|\psi_i\rangle=\delta_{mn}+i\Gamma^A_{mn}$,
the correlation matrix looks like $\langle\psi_i|d'_pd'_q|\psi_i\rangle=\delta_{pq}+ i\Gamma'^{B}_{pq}$, 
where $\Gamma'^B=W_f\Gamma^AW_f^{\dagger}$. The point to note here is that the matrix $\Gamma^A$ is calculated using the parameters 
of the initial Hamiltonian $H_i$ given by $\Gamma^A=W_i^{\dagger} \Gamma^B W_i$ where the 
expression of $\Gamma^B$ is shown in Eq.~(\ref{eq_gb}). Under the non-equilibrium dynamics, the time dependence of the above mentioned
correlation matrix  can be determined as $\langle\psi_i|d'_p(t)d'_q(t)|\psi_i\rangle=\delta_{pq}+ i\Gamma'^{B}_{pq}(t)$ where 
$\Gamma'^B(t)=\exp(i H_f t)\Gamma'^B \exp(-i H_f t)$.  
Finally, in order to calculate the time evolved EE, we compute the time-dependent correlation matrix  after a sudden quench 
 given by $\langle\psi_i|e_m(t)e_n(t)|\psi_i\rangle=\delta_{mn}+ i\Gamma^{A}_{mn}(t)$, where
\be
\biggr[\Gamma^A(t)\biggl]_{2l \times 2l} =
\biggl[ W_f^{\dagger} \biggr]_{2l\times 2N} \biggl[ \Gamma'^B(t) \biggr]_{2N \times 2N} \biggl[ W_f \biggr]_{2N\times 2l}.
\label{eq_gat}
\ee
For each time instant the matrix in Eq.~(\ref{eq_gat}) is diagonalized and then the EE can be calculated using 
Eq.~(\ref{eq_ee}) as a function of time.

\section{Results}
\label{result}

We here investigate the effect of the impurity term $H_{\rm imp}=-i\lambda_d a_1 a_2$ in the critical behavior of EE 
under both equilibrium and non-equilibrium cases where the $H_{\rm imp}$ is only applied to the final 
quenched Hamiltonian.

\subsection{Equilibrium}
\label{result_eq}
The quantum information theoretic measures such as the fidelity~\ct{gu10}, the Loschmidt echo~\ct{quan06,sharma12,rajak14u}, 
the quantum discord~\ct{sarandy09,dillenschneider08}, and 
the entanglement entropy are currently being studied intensively in the 
context of characterizing QPTs in various condensed matter systems.
In this section, we will first show the EE as an indicator of the QPTs of the model (\ref{ham1}).
We open our study by calculating the derivative of the EE, and plot as a function of $\la_2$
 to get the phase transition points as shown in Fig.~(\ref{ee_derivative1}). 
The derivative of EE shows dip, peak or kink at the QCPs. It is noteworthy that the impurity has 
a noticeable effect on the derivative $\Delta S/\Delta \la_2$
over the phase boundaries separating a topological or a non-topological phase from $n=2$ topological
phase, since it destroys two Majorana modes of the left end of the chain. 
In the inset of Fig.~(\ref{ee_derivative1}), we have plotted the EE as 
a function of $\la_2$ for various values of $\la_d$ by fixing $\la_1=1$. 
It shows that the peak value of the EE 
decreases at $n=2$-$n=1$ and $n=2$-$n=0$ phase boundaries with increasing $\la_d$, on the 
other hand, it indeed increases 
for $n=1$-$n=0$ boundary. This feature gets reflected in the derivative where 
the dip height in $n=2$-$n=1$ and $n=2$-$n=0$ phase boundaries reduces with $\la_d$ and peak height in 
$n=1$-$n=0$ boundary enhances.

\begin{figure}[ht]
\begin{center}
\includegraphics[height=5.4cm]{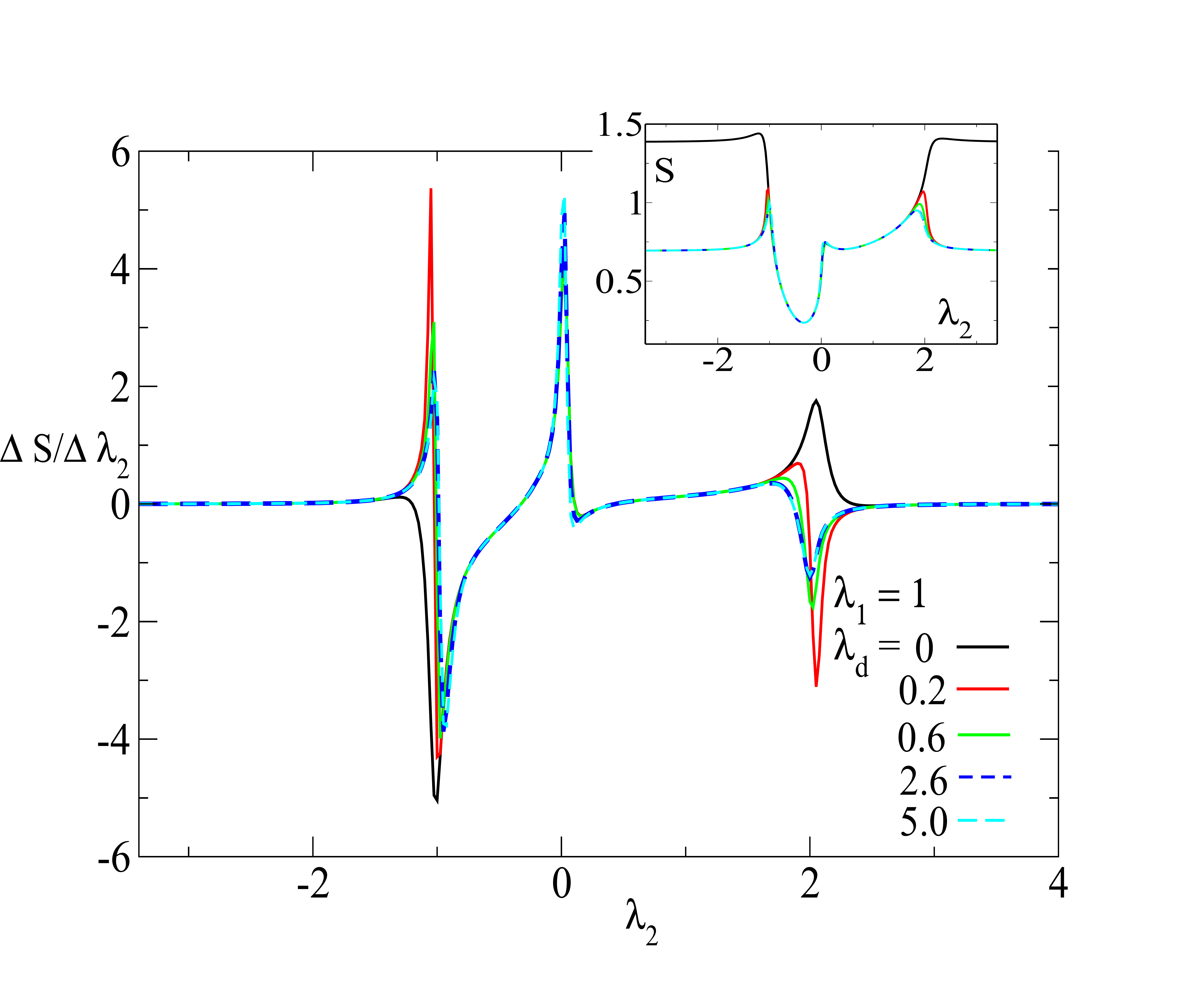}
\end{center}
\caption{(Color online) The derivative of EE with $\la_2$ shows peak, dip or kink at the phase boundaries pointing
towards the fact that it can be used as an indicator of QPTs. For $\la_d=0$, the derivative
shows a dip at $n=2$-$n=0$ phase boundary while the peaks are observed for $n=1$-$n=0$ and $n=1$-$n=2$ phase 
boundaries. These dip and peak in the derivative of EE that appear in the phase boundaries separating $n=2$ phase with others 
change to a kink like structure for finite $\la_d$. This qualitative change in behavior of the derivative is 
clearly evident from the variation of EE with $\la_2$ as shown in the inset (we set $\la_1=1$); the value of EE reduces inside the 
$n=2$ phase when $\la_d$ becomes finite.  
The dip/peak height of the derivative
at these kinks  is maximum for an infinitesimal impurity strength and decreases with increasing 
$\la_d$; inset suggests similar feature that at $n=2$-$n=0$ and $n=1$-$n=2$ phase boundaries 
the EE shows a peak that decreases with $\la_d$.
In contrast, at $n=0$-$n=1$ phase boundary, the peak height of the derivative enhances with $\la_d$ as 
peak height of EE increases.  
We choose the system size to be $N=100$ and block length to be $l=30$.}
\label{ee_derivative1}
\end{figure}

The noteworthy feature observed in the derivative at these phase boundary points is that the 
peak/ dip for $\la_d=0$ changes to dip/ peak for any finite $\la_d$ and the 
height associated with them decreases with increasing $\la_d$. These behavior are observed over the phase 
boundaries that separates $n=2$ phase from others. For, $\la_d=0$, 
the EE decreases when the system enters from $n=2$ phase to $n=0$ phase.
For finite $\la_d$, vice versa occurs. Hence, the dip observed in the derivative turns into a peak for a finite $\la_d$.  
On the other hand, the peak structure remains unaffected over the $n=0$-$n=1$ phase boundary as
the impurity term  is  introduced.

We shall now analyze the behavior of EE in three phases extensively as 
shown in the inset of fig.~(\ref{ee_derivative1}). For $\la_d=0$, the EE remains 
almost constant inside lower $n=2$ phase and then after it starts decreasing around $\la_2=-1$; it reaches  minimum 
value in the $n=0$ phase. Afterwards it  starts  increasing up to $\la_2=2$, in between, showing a kink around $n=0$-$n=1$ boundary (where $\la_2=0$).
Finally it saturates inside the upper $n=2$ phase to the same value as of the lower one.
Furthermore, the value of EE reduces in both the $n=2$ phases in the presence of the impurity, whereas it  
remains almost unchanged inside $n=0$ and $n=1$ phases. 

We can explain this phenomena qualitatively by considering that
the EE in Eq.~(\ref{eq_ee}) consists three contributions, given by
\be
S(l)=S_{BU}+S_B^L+S_B^R,
\label{ee_split}
\ee
where the contributions $S_{\rm BU}$, $S_B^L$ and $S_B^R$ come from bulk, left and right boundary modes
of the system respectively. The EE inside the $n=2$ phases has all three contributions of Eq.~(\ref{ee_split}), 
since there exists two Majorana modes in each end. On the other hand, the $n=0$ phase does not host
any Majorana edge mode that results only bulk contribution in the EE. Again, since the $n=1$ phase has one Majorana 
mode at each end, the EE has all three contributions but the value will be less than that of the $n=2$ phase. 
Now, once the impurity term is applied, the constant value of EE obtained for $\la_d=0$ case 
inside the $n=2$ phases decreases, but it does not depend on the impurity strength. 
This is due to the fact that the application of the impurity term, the 
left end Majorana modes of $n=2$ phases vanish, although the right end modes remain intact. 
As a result, the left end modes do not contribute in the EE (see Eq.~(\ref{ee_split})) that reduces 
the EE compared to the $\la_d=0$ case.
On the other hand, inside $n=0$ and $n=1$ phases the EE for $\la_d=0$ and $\neq0$ coincides 
with each other. This phenomena can be explained using the fact that the Majorana 
mode of $n=1$ phase remains unaffected by the impurity term. Therefore, the EE has all three 
terms as described in Eq.~(\ref{ee_split}) even after the application of $\la_d$. 
As mentioned already, since the $n=0$ phase does not have any zero mode it remains unaffected 
by the impurity. This explains the behavior of the EE inside the $n=0$ and $n=1$ phases in the 
presence of the impurity.

\begin{figure}[ht]
\begin{center}
\includegraphics[height=5.4cm]{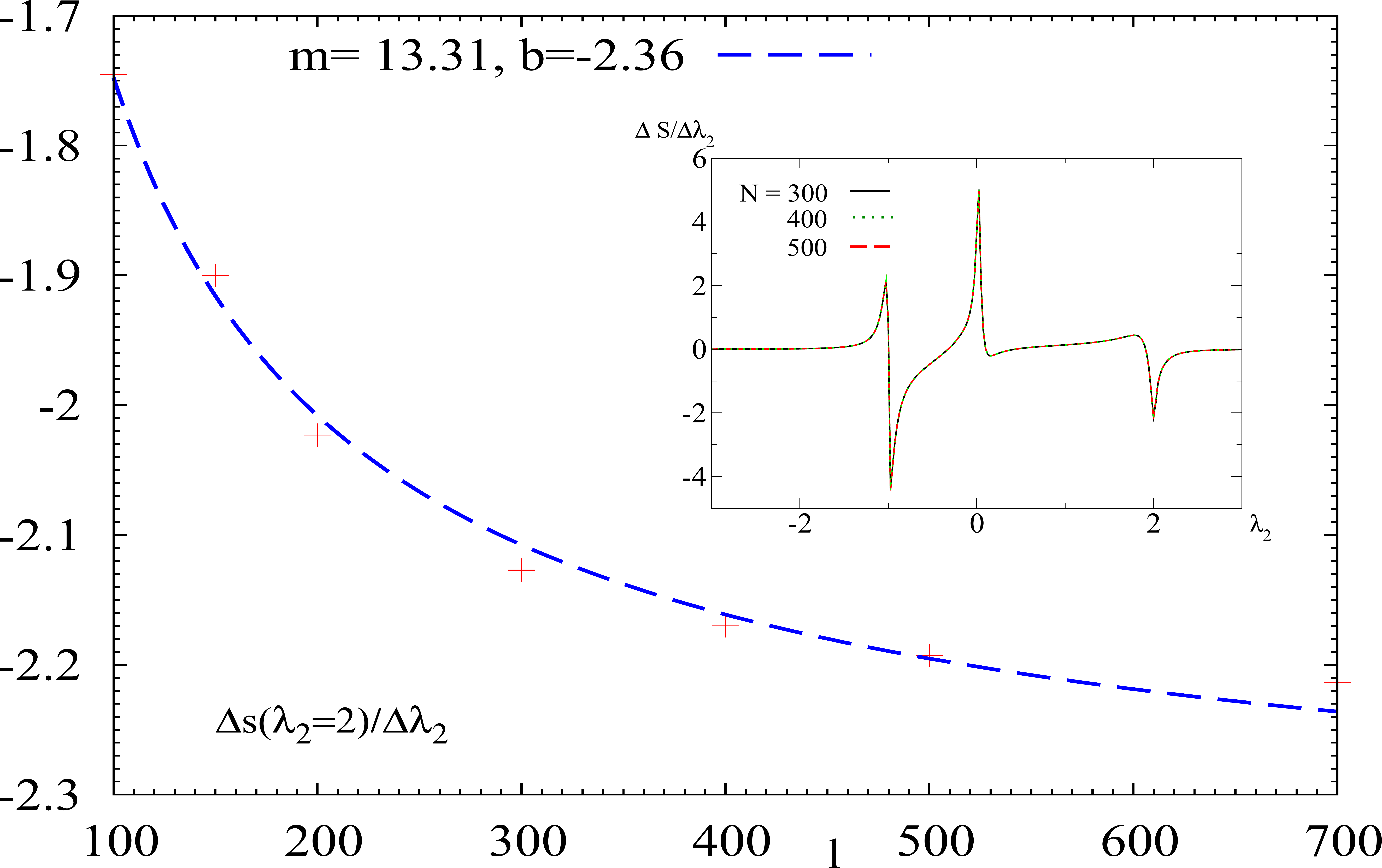}
\end{center}
\caption{(Color online){ Plot shows that peak/ dip,
observed in the derivative of EE for a clean system with $\la_d=0.1$, becomes more sharper  
with block size $l$. We choose $n=1$-$n=2$ phase boundary.
We see that $\frac{\Delta S(\la_2=2)}{\Delta \la_2}$ matches well 
with $ (m/l)\log l +b$ (depicted by blue dashed lines) where
$m$ and $b$ is found to be $13.31$ and $-2.36$.
Inset shows the variation of the derivative with $l$ by varying $\la_2$ and $\la_1$ is kept fixed at unity. }
}
\label{ee_derivative2}
\end{figure}

{In parallel, we investigate the height of the singularities in 
derivative of the EE as a function of block size $l$ for  $\la_d=0.1$. 
It can be noted that dip/ peak in Fig.~\ref{ee_derivative2}
becomes sharper with increasing block length $l$.
Here we have studied height of dips at $n=2$-$n=1$ boundary  and 
found that it 
varies as $l^{-1}\log l$. Our result for derivative of the EE is in accordance with the study of the 
finite size effect of EE in one dimensional topological system \ct{wang17}. 
Similar to scaling function associated with free energy \ct{gulden16}, here also
 the finite size scaling of EE is sensitive to the topological character of the model.}

\begin{figure}[ht]
\begin{center}
\includegraphics[height=6.7cm]{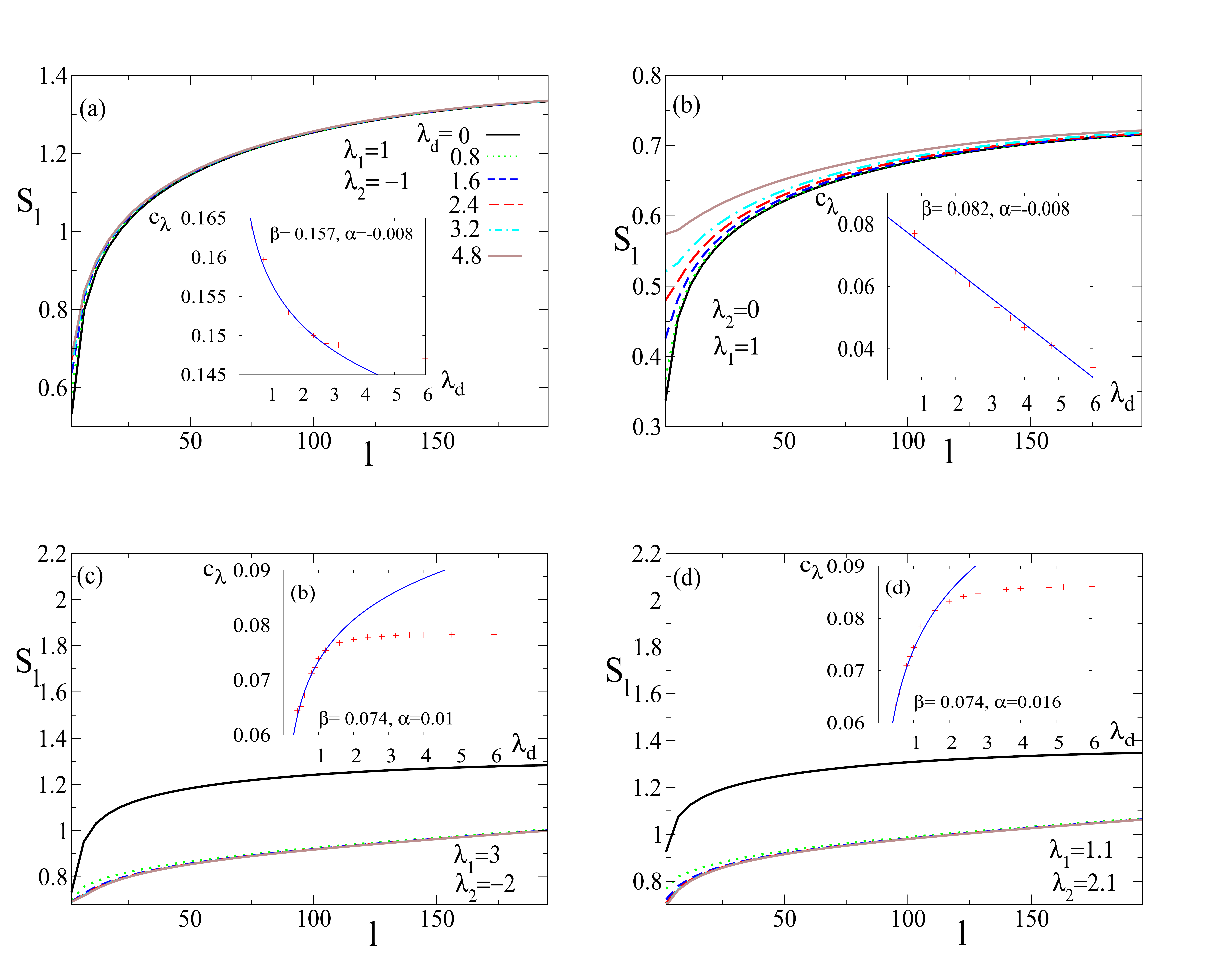}
\end{center}
\caption{(Color online) The equilibrium EE, $S_l$ as function of block length $l$ for different values 
of $\la_d$ is plotted for (a) $n=2$-$n=0$ (with $\la_2=-1$), (b) $n=1$-$n=0$ (with $\la_2=0$), (c) $n=2$-$n=1$ (with $\la_2=-2$)
and (d) $n=2$-$n=1$ (with $\la_2=2.1$) phase boundaries. 
{For (a) and (b), the EE becomes maximum for $ \la_d=4.8$ as shown by the solid
brown line appeared at the top of the plots. The 
dotted-long-dashed sky line  with $ \la_d=3.2$,  long-dashed red line with $ \la_d=2.4$,
short-dashed blue line with $ \la_d=1.6$, dotted green line with $ \la_d=0.8$, and 
solid black line with $\la_d=0$ appear in a 
decreasing order.
The order is reversed for (c) and (d) compared to (a) and (b).}
The value of EE with the impurity strength $\la_d$ increases for (a) and (b), whereas decreases for (c) and (d). 
The inset shows the 
variation of the effective central charge $c_{\la}$ as a function of $\la_d$. The behavior exhibited by 
$c_{\la}$ is opposite to that of the EE in all the above cases. Additionally, insets of (a), (c) and (d) show
that $c_{\la}$ varies logarithmically with $\la_d$ for $\la_d^l<\la_d<\la_d^u$;
$c_{\la}= \alpha \log \la_d +\beta$ (indicated by blue solid lines). 
On the other hand, for (b) it is linear throughout the range of
$\la_d$; $c_{\la} = \alpha \la_d+\beta$ (drawn using blue solid lines). In this case, $\beta$ equals to the central charge ($c_0=1/12$) 
for the clean system. Numerically, the extrapolated value of $c_\la$ at $\la_d=0$ is $\beta=c_0=0.082$
(close to $1/12$).}
\label{ee_equi}
\end{figure}

Our aim is now to study the variation of $S_l$ with $l$ for different values of impurity strength over 
various phase boundaries (see Fig.~\ref{ee_equi}). As mentioned before, 
the critical EE shows a scaling relation $S_l=c_0 \log l+\gamma$ with the block size $l$, 
where $c_0=c/6$  and $c$ being the central charge and $\gamma$ 
is a non-universal constant. 
As shown in Fig.~\ref{ee_equi}, 
in the presence of the impurity term the EE follows
similar scaling relation as in the clean case with an effective $c_0$ (namely, $c_{\la}$) 
and non-universal constant $\gamma_{\la}$ which depend on $\la_d$: $S_l=c_{\la} \log l+\gamma_{\la}$.
The EE on the anisotropic critical line $n=0$-$n=2$ is minimally increased by 
$\la_d$ (see Fig.~\ref{ee_equi}(a)), whereas
the  EE increases substantially for Ising critical line with $\la_2=0$ (see Fig.~\ref{ee_equi}(b)). 
In contrast, for $n=1$-$n=2$ phase boundaries the value of EE reduces considerably compared to 
$\la_d=0$ once a finite $\la_d$ is applied (see Fig.~\ref{ee_equi}(c) and (d)). However, it 
does not change so much for two different values of $\la_d$. 
We have plotted $c_{\la}$, which essentially captures the signatures of the effective central 
charge,  as a function of $\la_d$ for all cases in the insets 
of Fig.~\ref{ee_equi}. Interestingly, for all phase boundaries the central charge as a function of 
$\la_d$ exhibits exactly an opposite behavior as compared to the EE in terms of decreasing or increasing nature.
It seems that these two behaviors  contradictory to each other i.e., when 
EE decreases with $\la_d$, central charge increases.
However, this is indeed easy to explain 
using $\gamma_{\la}$ which also changes with $\la_d$ in an opposite way compared to $c_{\la}$.

\begin{figure}[ht]
\begin{center}
\includegraphics[width=4.20cm,height=3.50cm]{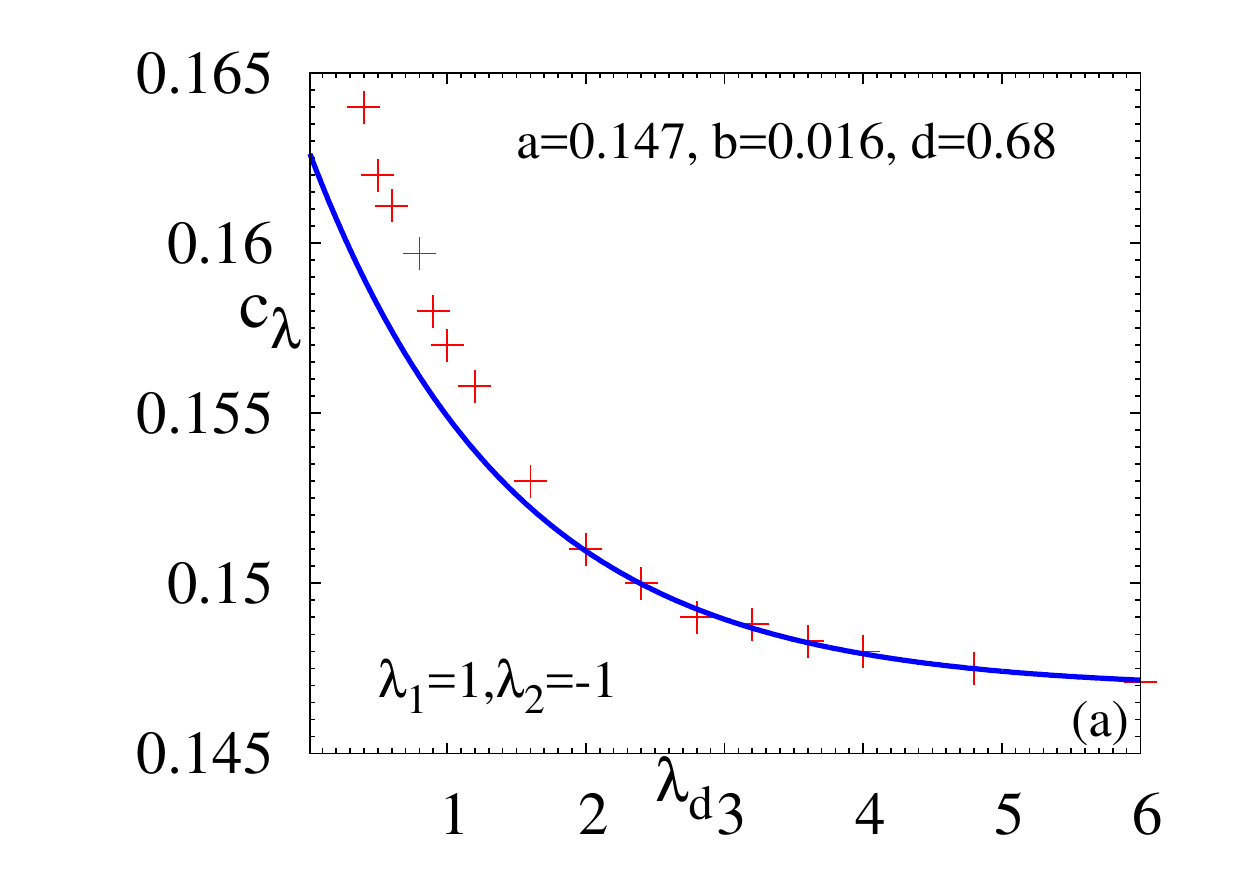}
\includegraphics[width=4.20cm,height=3.50cm]{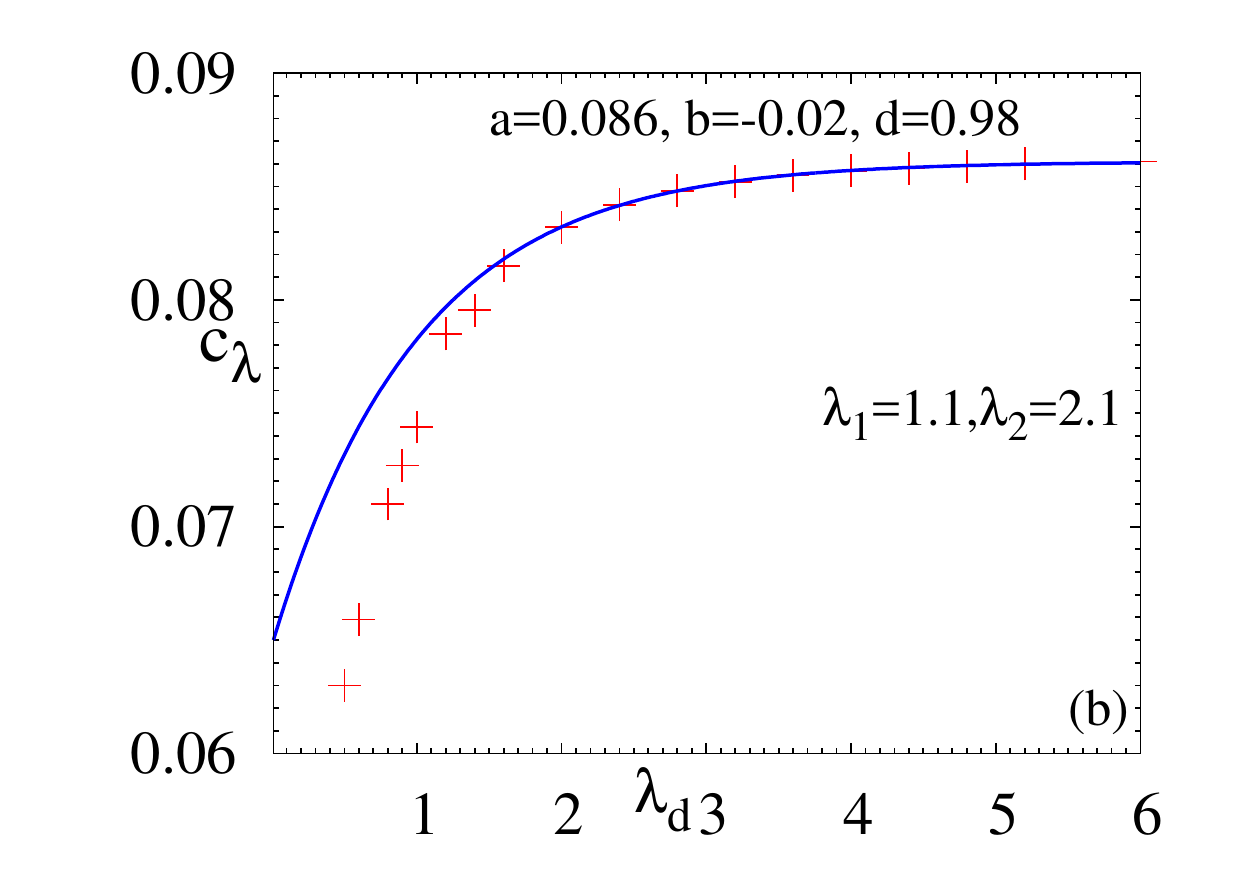}
\end{center}
\caption{(Color online) Plots (a) and (b) show the variation of $c_{\la}$ with 
$\la_d$ as shown in Fig.~\ref{ee_equi} (a) and (d) 
respectively. Here, it has been seen that both the plots can be fitted with $c_{\la}=a+b\exp(-d\la_d)$ 
for $\la_d>\la_d^u$ as shown by blue solid lines. It indeed indicates that the effective central charge eventually saturates for 
strong impurity limit with an exponential damping. The values of fitting parameters are provided inside the plots.}
\label{ee_equi_fit}
\end{figure}

Let us now extensively investigate the behavior of central charge as reflected in $c_{\la}$ with $\la_d$. 
A clean spin chain (i.e., with $\la_d=0$) with open boundary condition in one dimension has the central charge
$c=1/2$ and $1$ (with $c_0=1/12$ and $c_0=1/6$) for Ising and anisotropic critical lines respectively.
We note that for Figs.~\ref{ee_equi}(b), (c) and (d) with $\la_d=0$, $c_0$ close to $1/12$ which correspond to Ising critical line, 
whereas for Fig.~\ref{ee_equi}(a), $c_0\sim 0.161$ that represents the anisotropic critical line.

For the phase boundaries separating $n=2$ phase from others (see inset of Fig.~(\ref{ee_equi}) (a), (c),
and (d)), a careful analysis suggests that effective central charge contained in $c_{\la}$ that follows the relation:
$c_{\la}\approx\alpha \log(\lambda_d) +\beta$ for a range $\la_d^l<\la_d<\la_d^u$, where 
the values of $\la^l_d$ and $\la^u_d$ depend on the strengths of $\la_1$ and $\la_2$.
Afterwards, it eventually saturates with an additional exponential damping term
given by $\exp(-\la_d)$ (see Fig.~\ref{ee_equi_fit}). The saturation characteristics of $c_{\la}$ for strong 
impurity limit can be explained by the mathematical form given by $c_{\la}=a+b\exp(-d\la_d)$; hence $\la_d \to \infty$, $c_{\la} \to a$.

In contrary, on the $n=1$-$n=0$ phase boundary (with $\la_2 = 0$) the effective central charge shows 
a linear relation with $\la_d$, $c_{\la}=\beta+\alpha \lambda_d$. Interestingly, in this case, 
the value of $\beta$ is close to $c_0=1/12$ which is the central charge on that boundary 
with $\la_d=0$. Therefore, the term $\alpha\la_d$ can be considered as the correction over 
the bare central charge due to the application of the impurity.
It is noteworthy that over the complete phase boundary with $\la_2 \neq 0$ and even in the strong impurity limit 
this linear behavior remains unaffected. 
Hence it can be inferred that the impurity term indeed 
plays a distinctly different role in the phase boundaries shared with $n=2$ phase compared to others.

In this connection, we would like to mention that for a 
 disordered quantum spin chain, an effective  central charge comes into play instead of the bare
 central charge obtained in the 
clean limit \ct{refael07}. We here show that even a single impurity term can also lead to an effective central charge. 
At the same time, this effective central charge has distinct scaling relations with the strength 
of the impurity over different phase boundaries.

\subsection{Non-equilibrium}
\label{result_non_eq}

In the previous section, we have discussed the influence of the impurity term on both critical and 
off-critical EE when the system is in equilibrium.
Provided the  general formalism for calculating the time evolution of the EE 
with a complex term in the Hamiltonian in Eq.~(\ref{ham2}) as presented in Sec.~\ref{EE},
here we shall now investigate the effect of the impurity term $H_{\rm imp}=-i\lambda_d a_1 a_2$ on the 
evolution of the EE when the Majorana chain is suddenly quenched to various critical points.

\begin{figure}[ht]
\begin{center}
\includegraphics[width=6.0cm]{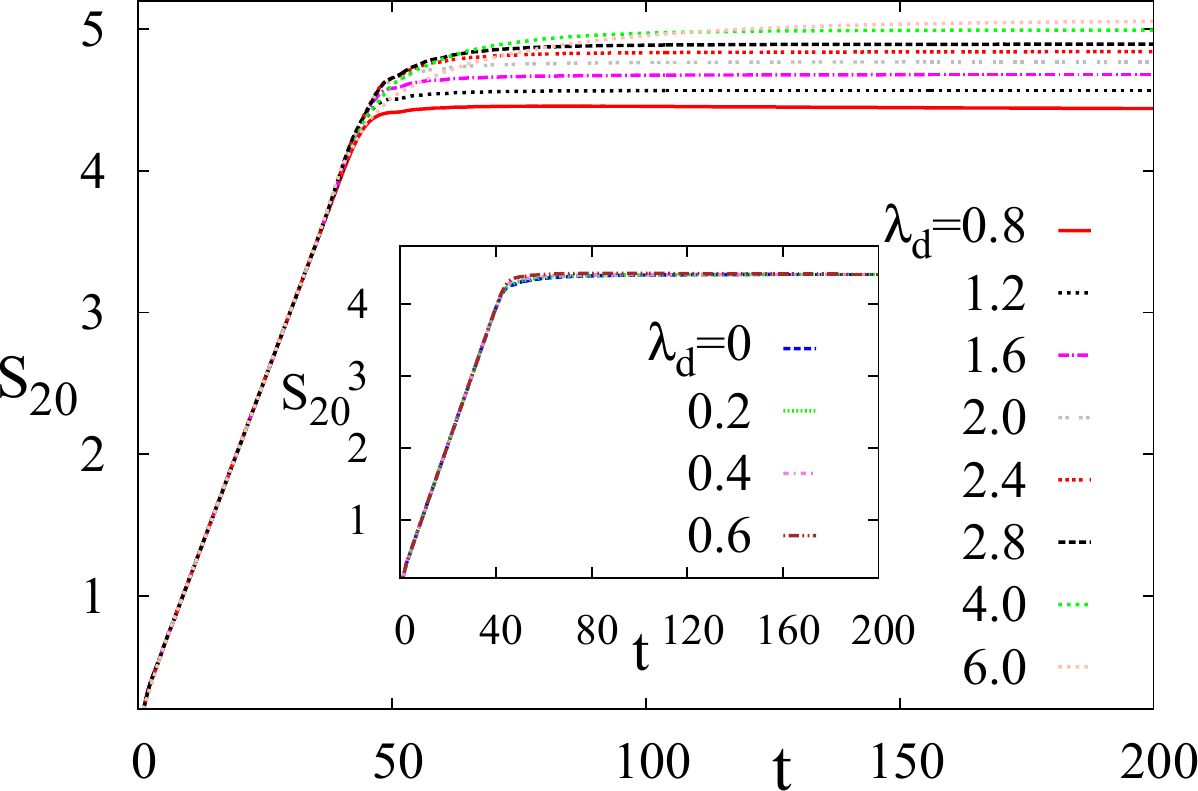}
\end{center}
\caption{(Color online) Evolution of the EE as a function of time for different values of 
the impurity strength ($\la_d$). It can be observed that the EE gets affected by the $\la_d$ after a 
certain value of $\la_d$
when the system is quenched to $n=0$-$n=1$ phase boundary (with $\la_1=1$ and $\la_2=0$) 
from $n=0$ phase with $\la_1=0.5$. 
{ The different curves 
from bottom to top are given by, solid red line ($\la_d=0.8$), 
dashed black line ($\la_d=1.2$), long-dashed-dotted pink line ($\la_d=1.6$), 
double dotted grey line ($\la_d=2.0$), dashed red line ($\la_d=2.4$),
long-dashed black line ($\la_d=2.8$), dashed green line ($\la_d=4.0$), dashed orange line ($\la_d=6.0$).}
Inset: It shows that saturation value of the EE is almost 
independent $\la_d$ upto some value, say $\la_d^*$. At the same time, main plot depicts that 
the saturation value of the EE increases logarithmically with $\la_d$ after $\la_d^*$.
Here, block length, $l=20$ and $N=300$. }
\label{ee1}
\end{figure}

Let us first assume a situation where $\lambda_1$ is suddenly changed  
from $n=0$ phase to $n=1-n=0$ phase boundary by fixing $\lambda_2$ at $0$. The EE after the sudden quench 
increases linearly up to a time $t^*=l/v$ where $v$ is the group 
velocity of the quasiparticles generated due to the sudden quench (see Fig.~\ref{ee1}) 
and $l$ being the length of subsystem. This phenomena has been explained in the 
earlier related literature
using the picture of quasiparticle propagation through the system 
after the  quench \cite{calabrese04}.
We find that the group velocity $v$, numerically calculated from the final real space 
Hamiltonian, is almost independent of $\lambda_d$ and 
it remains at nearly equal to $0.5$ in this case. This also can be 
observed from Fig.~\ref{ee1} where the EE for a block of length $l=20$ shows a linear rise with time up to 
$t^*\simeq 40$ with different values of $\la_d$. After time $t=t^*$, the EE saturates at some
finite values that depend on the strength of $\la_d$. The inset of Fig.~\ref{ee1} shows that the EE curves 
almost overlap with each other even in the saturation region up to a threshold value of 
$\la_d$, denoted by $\la_d^*$. At the same time, for $\la_d>\la_d^*$, the saturation value increases 
with $\la_d$.
It has been observed from the plot 
that the value of $\la_d^*$ depends only on the final values of parameters $\la_1$ and $\la_2$.

One can see from Fig.~\ref{ee2} that the linear behavior
of the EE persists up to $t=l$ for the $n=0-n=2$ boundary as $v$ becomes unity there. 
In contrast to the previous case, here the saturation value of 
the EE indeed decreases by a small amount with increasing $\la_d$ when $\la_d<\la_d^*$.
Here also the saturation value of the EE increases with $\la_d$ after $\la_d>\la_d^*$.
On the other hand, the EE increases linearly up to $t=2l/3$ as $v\simeq 3/2$ over the $n=2-n=1$ phase boundary 
(see the Fig.~(\ref{ee3})). In this case, the decrease of the saturation value of the EE with 
increasing $\la_d$ (up to $\la_d^*$) is more prominent than the $n=0-n=2$ boundary.
However, the behavior of the EE for $\la_d>\la_d^*$ seems to be similar to the previous cases.

\begin{figure}[ht]
\begin{center}
\includegraphics[width=6.0cm]{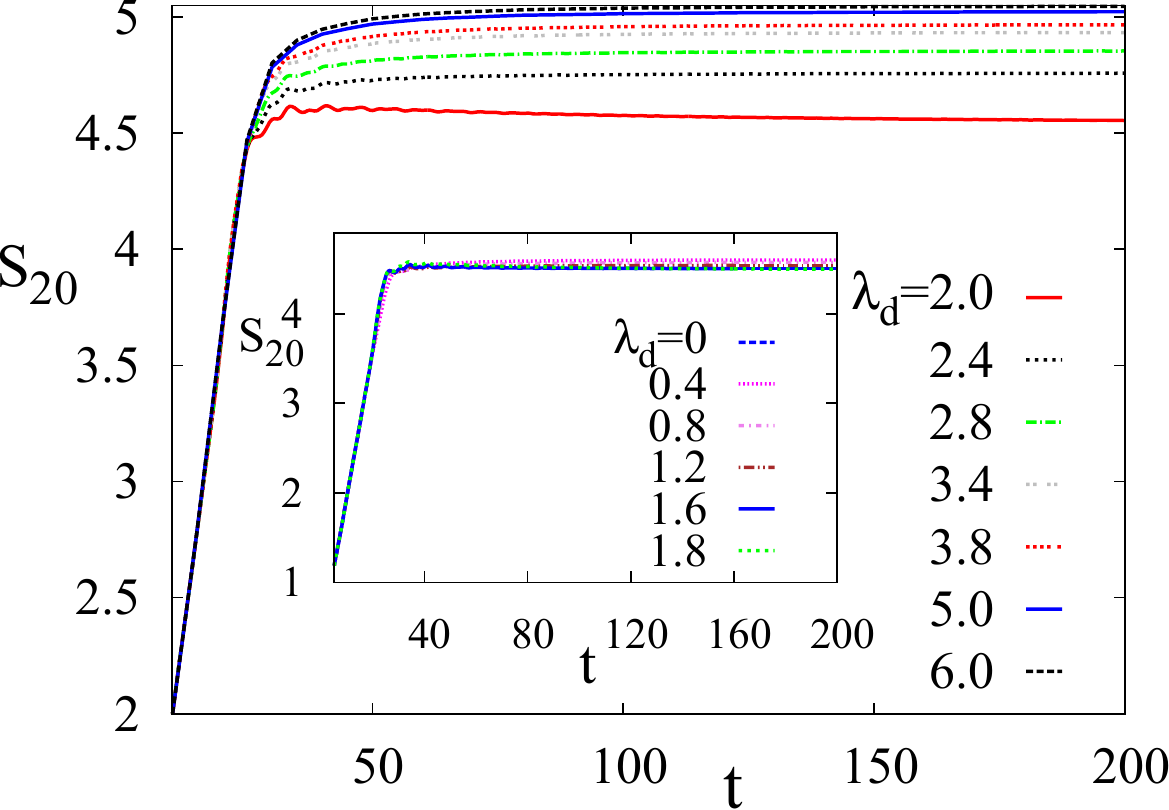}
\end{center}
\caption{(Color online) Variation of the EE with time after quenching the system  
to $n=0$-$n=2$ boundary ($\la_1=1$ and $\la_2=-1$) from $n=0$ phase with $\la_2=-0.5$ and $\la_1=1$. 
{ The various curves
from bottom to top are given as solid red line ($\la_d=2.0$), 
dashed black line ($\la_d=2.4$), long-dashed-dotted green line ($\la_d=2.8$), 
double dotted grey line ($\la_d=3.4$), dashed red line ($\la_d=3.8$),
solid blue line ($\la_d=5.0$), long-dashed black line ($\la_d=6.0$). }
Here, block length, $l=20$ and $N=300$. Inset plot shows that saturation value of the EE is 
weakly dependent on $\la_d$ and decreases by a small amount with increasing the value of $\la_d$ upto 
a certain value, say $\la_d^*$. On the other hand, the main plot depicts that  
the saturation value of EE increases with $\la_d$ in a non-linear fashion after $\la_d^*$.}
\label{ee2}
\end{figure}

We are now interested to determine the relation between the saturation value of 
the EE and $\la_d$. It can be observed from Fig.~\ref{ee4} that 
the variation of the saturation value of EE with $\la_d$ ($>\la_d^*$) is given by
$S_{\rm sat}\propto \log \lambda_d$ for all three cases discussed above.
The logarithmic behavior of EE  suggests 
the fact that $\la_d$ affects the EE in an identical manner irrespective 
of the nature of the phase boundary i.e., whether the phase boundary separates a topological 
phase from a non-topological phase or two different topological phases.
The semiclassical theory of the EE \ct{rieger11} suggests that the more number of 
quasiparticles is generated as one increases the strength of impurity 
and as a result saturation value of the EE 
increases with $\lambda_d$. However, the logarithmic dependence of saturation value of EE 
for $\la_d>\la_d^*$ can not be explained by this theory of quasiparticle generation.

Similar to the variation of $c_{\la}$ on the phase boundary shared with $n=2$ phase
as shown in Fig.~(\ref{ee_equi_fit}), we find that the saturation 
value of EE eventually approaches to a fixed value with $\la_d$ for strong impurity limit (see Fig.~\ref{ee5} (a)). 
In contrary to the equilibrium case, the saturation in $S_{\rm sat}$ is also observed 
for $n=1$-$n=0$ boundary (see Fig.~\ref{ee5} (b)), whereas 
$c_{\la} $ shows linear variation for whole range of $\la_d$ there. 
The strong impurity limit here is meant to be above the range of $\la_d$ within which 
logarithmic rise of $S_{\rm sat}$ is observed. As described earlier, it might be the case 
that after a cut-off value of $\la_d$ the rate of
quasiparticle generation decreases with an exponential damping factor
resulting $S_{\rm sat}=a+b\exp(-\la_d)$
for strong impurity limit. 
It can be noted that the ranges of $\la_d$ within which
the logarithmic rise in $c_{\la}$ and $S_{\rm sat}$ occurs following saturation
are different from each other.

\begin{figure}[ht]
\begin{center}
\includegraphics[width=6.0cm]{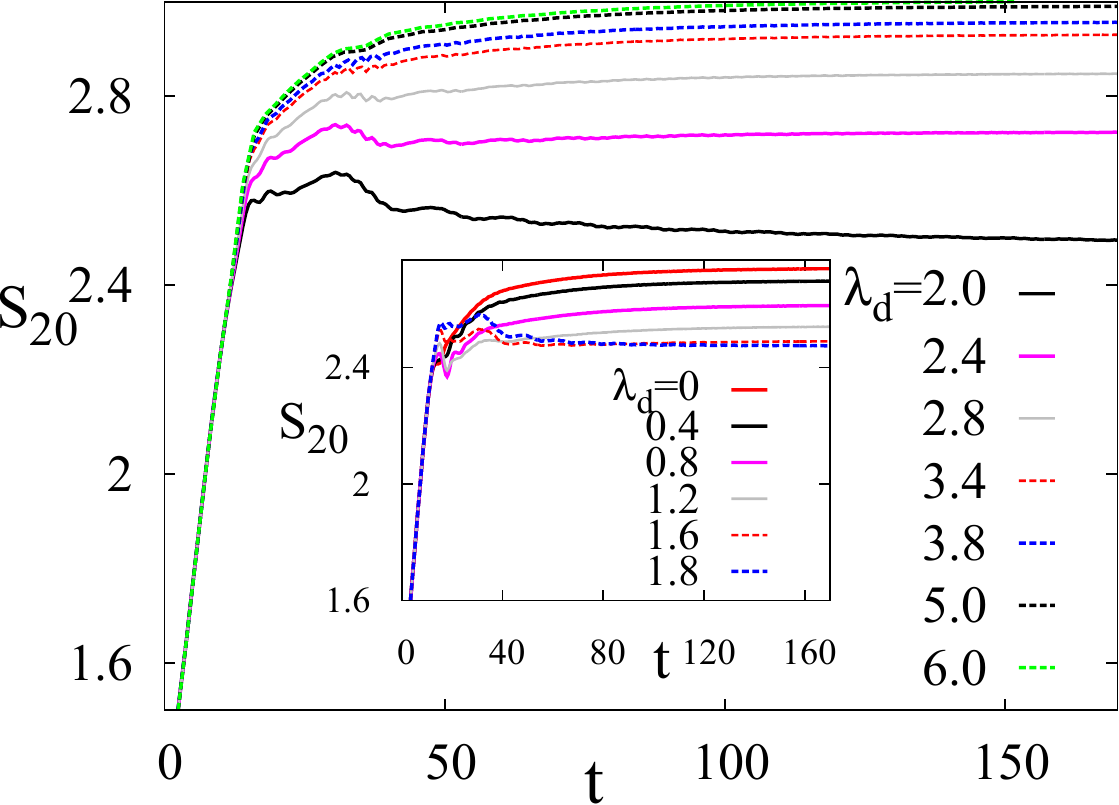}
\end{center}
\caption{(Color online) Time-evolution of the EE 
is affected by the impurity strength as the system is quenched 
to $n=1$-$n=2$ boundary (with $\la_1=1$ and $\la_2=2$) from $n=2$ phase with $\la_1=0.5$ and $\la_2=2$. 
{ The different curves
from bottom to top are denoted by solid black line ($\la_d=2.0$), 
solid pink line ($\la_d=2.4$), solid grey line ($\la_d=2.8$), 
long-dashed red line  ($\la_d=3.4$), long-dashed blue line ($\la_d=3.8$),
long-dashed black line ($\la_d=5.0$), long-dashed green line ($\la_d=6.0$). }
Here, block length, $l=20$ and $N=300$. 
Inset: It can be seen that the saturation value of the EE 
decreases slowly when $\la_d$ increases up to a certain value $\la_d^*$.
Main plot: Similar to the previous cases, after $\la_d^*$, the saturation value of the EE
increases with $\la_d$ in a non-linear fashion.}
\label{ee3}
\end{figure}

\begin{figure}[ht]
\begin{center}
\includegraphics[width=2.5in]{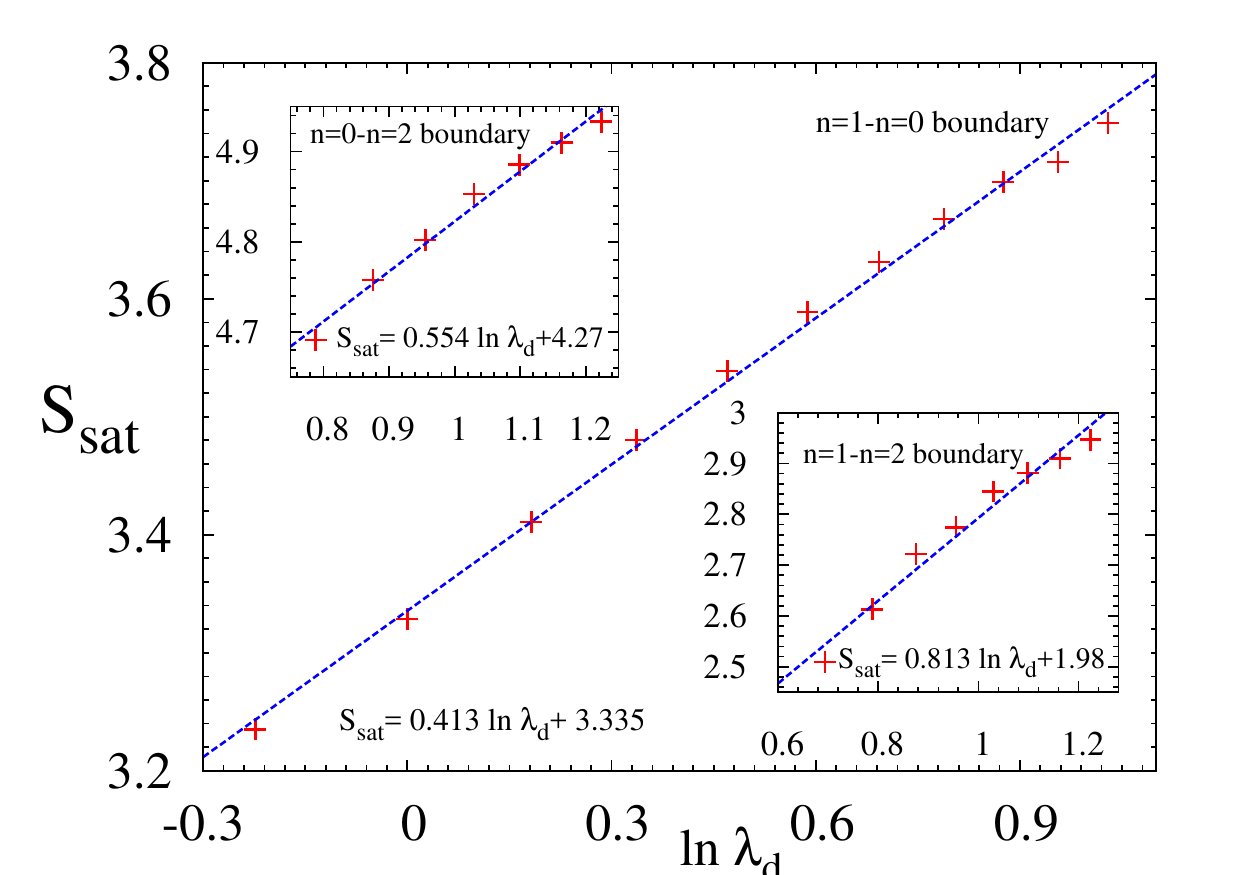}
\end{center}
\caption{(Color online)
We have plotted the saturation value of the EE as a function of $\ln \la_d$ for three different 
cases studied in Figs.~\ref{ee1}, \ref{ee2} and \ref{ee3}. 
This plot shows that saturation value of the EE increases linearly with $\ln \la_d$ after $\la_d^*$. 
The blue dashed lines represent the fitted curve: $S_{\rm sat}=\alpha \ln \la_d+ \beta$.}
\label{ee4}
\end{figure}

\begin{figure}[ht]
\begin{center}
\includegraphics[width=3.60in,height=1.80in]{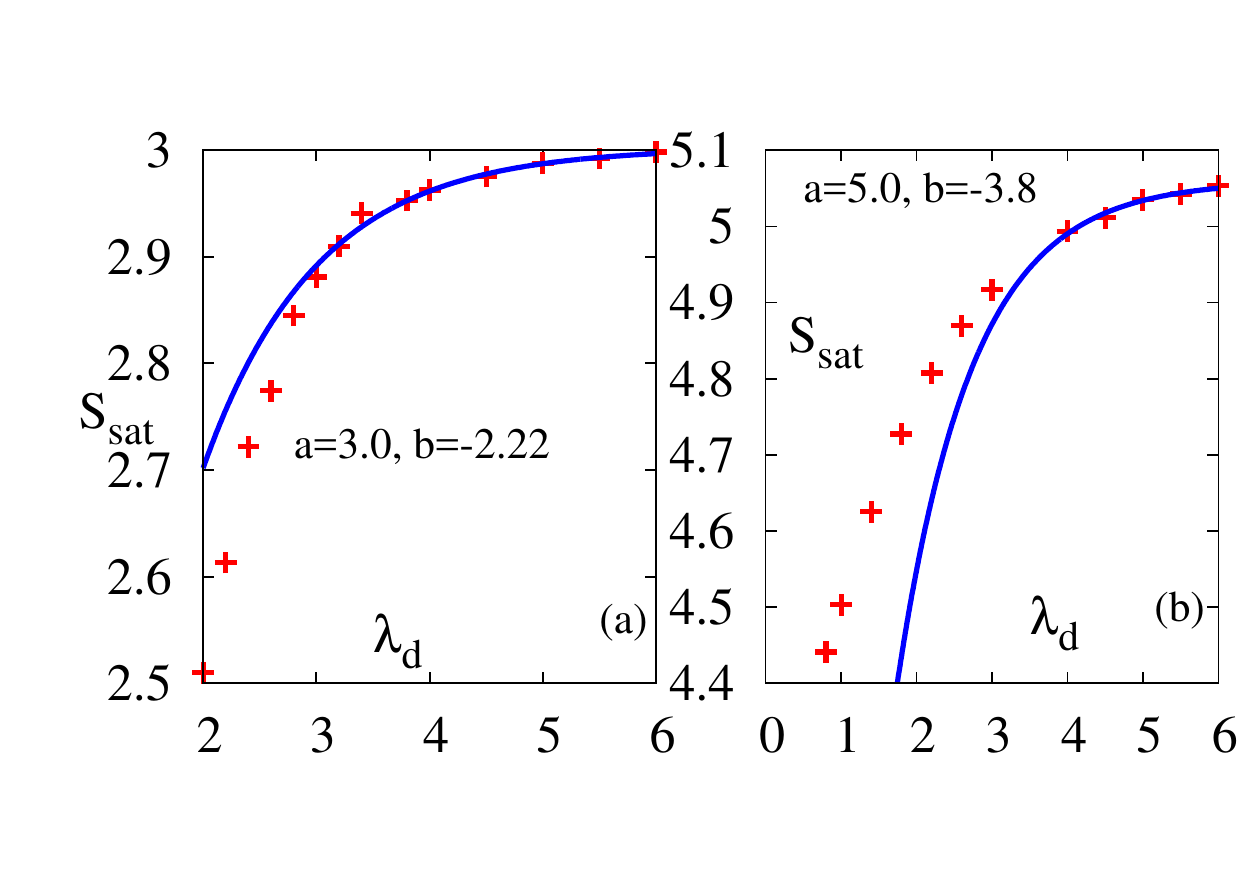}
\end{center}
\caption{(Color online) Variation of $S_{\rm sat}$ as a function of $\la_d$ when the 
quenching is performed up to (a) $n=1$-$n=2$ and (b) $n=1$-$n=0$ phase boundaries. 
After a logarithmic increase, $S_{\rm sat}$ follows the relation $c_{\la}=a+b\exp(-\la_d)$ (depicted by blue solid lines).}
\label{ee5}
\end{figure}

For a global quench to a critical point it has been shown that the EE follows the relations 
$S_l \sim c~t$ for $t< l/v$ and $S_l \sim c~l$
for $t>l/v$ \ct{calabrese05}; therefore, the central charge plays a crucial role in both the temporal regions.
In our case, interestingly, the linear rise of EE with $t$ remains almost unaltered even if $c_{\la}$ (i.e.,
essentially the effective central charge) depends on $\la_d$, although, some minor changes occur in a small time window where the linear rise terminates and the 
saturation starts. On the other hand, 
saturation values of EE behave in an identical manner as exhibited by $c_{\la}$ for boundaries shared with $n=2$ 
phase. Therefore, the upshot of 
$\la_d$ on the effective central charge gets imprinted on the saturation characteristics of EE. 
Considering the complete non-equilibrium evolution over all the phase boundaries,
one can say that the outcome of the 
effective central charge seems to behave differently with $\la_d$ as compared to the static limit.

In the present context, one can easily note the difference between equilibrium and non-equilibrium
scenarios. The non-equilibrium case which we consider can be illustrated as two
simultaneous quenches comprising of a global and a local quench. The global quench is performed
by changing a parameter of the Hamiltonian up to a critical point, whereas addition of an
impurity term at one end of the chain can be  considered as a local quench. Therefore
the behaviour of the EE with time is determined by both the quenches unlike the situation
to the equilibrium case where one impurity term is added in the critical chain. This might be
one of the reasons why the linear rise of the EE with time is not noticeably affected by the impurity term.  
In other words, for linear rise of the EE global quench dominates and effectively central charge
remains unaffected by the impurity term.
In addition, this anomalous behavior may be due to the fact that the EE in static limit is 
governed by low energy properties of the ground state only, whereas, due to the sudden quench, the behavior 
of EE in non-equilibrium case is substantially determined by the excited energy levels.
In this connection, we would like to mention that 
the contribution from higher excited state is extensively 
studied using the spectral function following a sudden quench between two different phases in the 
Lipkin-Meshkov-Glick model \ct{campbell16}.

\section{Conclusions}
\label{conclusion}
We investigate the critical characteristics of EE in both equilibrium and 
non-equilibrium situations by considering the effect of the impurity  in a three spin interacting model. We show that 
 the topological phase transitions can be detected by the derivative of EE that shows   
cusps in the vicinity of the phase boundaries. By applying the impurity term we can probe that
the edge modes do contribute in the EE. Additionally, our study suggests that the equilibrium 
EE satisfies a finite size scaling relation $l^{-1}\ln l$.
Interestingly, similar to disordered systems, the application of a single impurity in the system 
leads to the effective central charge while keeping the critical scaling relation (i.e., $\log l$) of the EE unchanged.
For phase boundaries connected with $n=2$ phase,
we find that the effective central charge shows a logarithmic scaling relation with $\la_d$
in a certain range of $\la_d$
 following the saturation with an exponentially damping factor at large $\la_d$.
At the same time, the central charge acquires a linear correction as a function of $\la_d$ 
over the bare value at the phase boundary separating $n=0$ and $n=1$ phases.

Furthermore, we extend our study 
to the time-evolution of the EE following the a critical quench where the impurity term is only added 
to the boundary of the quenched critical chain.
In this case, we focus on the effect of the impurity Hamiltonian 
on the saturation value of the EE as the linear rise remains unaffected.
Our study indicates  that irrespective of the phase boundaries there exists a threshold value of the impurity strength after 
which the saturation value of the EE increases logarithmically with impurity strength. 
The threshold value of the impurity depends on the final parameters of the critical Hamiltonian. 
In the strong impurity limit, the increase of saturation value of EE with $\la_d$ is suppressed by 
an exponential damping factor.

It has been observed that the effect of the impurity term shows up differently in the EE under
equilibrium and non-equilibrium situations i.e., change in the effective central charge due to impurity,
as probed in the equilibrium analysis of EE, is 
substantially visible in the later temporal saturation of EE not in the initial rise with time.
{We provide two probable reasons for this anomalous behavior. One is related 
to the competitive effects of two simultaneous quenches.}
The another one is due to the fact  that
the higher excited energy levels contribute in the dynamics of EE, whereas, equilibrium behavior of EE is completely
governed by the ground state. 
However, in both the equilibrium and non-equilibrium cases, the range of $\la_d$ within which 
the effects are appreciably visible depends on the phase boundary and the values of the other parameters.

In recent years entanglement entropy serves as an indicator of 
thermalization and many body localization. We would like to mention a few commnents in that 
direction. General belief 
says that the ballistic growth (i.e, linear in time) of EE is a signature of thermalization for non-integrable systems \ct{kim13}.
Although, in our case the impurity term, being quadratic in fermionic operator, is not able to
break the integrability 
of the system, we find a  linear rise of EE followed by a saturation. 
On the other hand, logarithmic growth of EE for many body localized
state is clearly distinguished from the dynamical evolution of EE in
thermalized phase \ct{nanduri14}.

\begin{acknowledgments}
The authors are grateful to Amit Dutta and Bikas K. Chakrabarti for enlightening discussion. 
TN and AR thank IIT Kanpur and SINP Kolkata respectively where the initiation of the work was done.
TN acknowledges Kush Saha for critically reading the manuscript.
A.R. acknowledges financial support from the Israeli Science Foundation Grant No. 1542/14.
\end{acknowledgments}

\appendix
\section{Entanglement entropy}
\label{appendixa}
To calculate the EE in Eq.~(\ref{ee_def}), we first have to determine 
the reduced density matrix $\rho_l$. The non-local nature of the underlying Jordan-Wigner fermion allows us 
to  construct a $2l\times 2l$ correlation matrix, given by
\be
\biggr[\Gamma^A_l\biggl]_{2l \times 2l} =
\biggl[ W^{\dagger} \biggr]_{2l\times 2N} \biggl[ \Gamma^B \biggr]_{2N \times 2N} \biggl[ W \biggr]_{2N\times 2l}.
\label{eq_ga}
\ee
The matrix $\Gamma^A_l$ is a skew-symmetric matrix which can be represented into the block-diagonal form $\Gamma^C_l=V\Gamma^A_lV^{\dagger}$ by an 
orthogonal transformation with $V$. Then the matrix $\Gamma^C_l$ 
can be written as
\be
\Gamma^C_l = \bigoplus_{j=1}^{l} \left[
\begin{array}{cc}
0 & \eta_j \\
-\eta_j &0
\end{array}
\right].
\label{eq_gamacl}
\ee
This defines a new set of Majorana fermion operators 
\be
c_{p} = \sum_{q=1}^{2l}V_{pq} e_q.
\label{cp_Hi}
\ee
In this basis, the new correlation matrix is given by
\be
\langle c_p c_q \rangle=\delta_{pq}+ i(\Gamma^{C}_l)_{pq}.
\ee
The Eq.~(\ref{eq_gamacl}) indicates that the $c$ Majorana fermions are correlated 
when their site indices are separated by $1$. We use this fact in our next steps 
of calculations.

Finally, we express the Majorana fermions in terms of usual complex 
fermions. We define $l$ fermionic modes from $2l$ Majorana operators
\be
f_j=\frac{c_{2j-1}+ic_{2j}}{2}.
\ee
By definition the fermionic modes satisfy the relations
\be
\langle f_jf_k\rangle = 0, \hspace{2mm} \langle f_j^{\dagger}f_k\rangle=\delta_{jk}
\frac{1+\eta_j}{2}.
\ee
It signifies that there has no correlation among the $l$ fermionic modes. 
Using this fact the density matrix of the $l$ fermionic modes can be written as 
a direct product of $l$ uncorrelated modes $\rho_l=\otimes_{n=1}^{l}\varrho_n$ with 
each $\varrho_n$ having eigenvalues 
$(1\pm\eta_n)/2$. Now, from definition of the EE in Eq.~(\ref{ee_def}), it is given 
by
\be
S(l)= -\sum_{n=1}^l \big[\big({1+\eta_n \over 2}\big)\log \big({1+\eta_n \over 2}\big) 
+\big({1-\eta_n \over 2}\big)\log \big({1-\eta_n \over 2}\big)\big].
\label{eq_eel}
\ee

\end{document}